\documentclass{article}
\usepackage{smc}
\usepackage{times}
\usepackage{ifpdf}
\usepackage[english]{babel}
\usepackage{cite}
\usepackage{tipa}
\usepackage{subfig}
\def\BibTeX{{\rm B\kern-.05em{\sc i\kern-.025em b}\kern-.08em
    T\kern-.1667em\lower.7ex\hbox{E}\kern-.125emX}}
   
\usepackage{dcolumn,booktabs}
\newcolumntype{d}[1]{D..{#1}}
\newcommand\mc[1]{\multicolumn{1}{c}{#1}} 

\usepackage{array}
\newcolumntype{P}[1]{>{\centering\arraybackslash}p{#1}}

\usepackage[table]{xcolor}

\usepackage{tikz}
\usetikzlibrary{arrows.meta}
\tikzset{>={Latex[width=2mm,length=2mm]},
        base/.style = {rectangle, rounded corners, draw=black,
                       minimum width=4cm, minimum height=1cm,
                       text centered, font=\rmfamily},
         blue/.style = {base, fill=blue!30},
         red/.style = {base, fill=red!30},
         green/.style = {base, fill=green!30},
         orange/.style = {base, minimum width=2.5cm, fill=orange!15,font=\rmfamily},
        }

\usepackage{amsmath, amssymb, latexsym}
\DeclareMathOperator{\ReLU}{ReLU}

\definecolor{fc}{HTML}{AF7AC5}
\definecolor{h}{HTML}{228B22}
\definecolor{bias}{HTML}{87CEFA}
\definecolor{noise}{HTML}{8B008B}
\definecolor{convd}{HTML}{5499C7} 
\definecolor{convs}{HTML}{64BBF5} 
\definecolor{convf}{HTML}{52BE80} 
\definecolor{pool}{HTML}{E78F33} 
\definecolor{up}{HTML}{52BE80}
\definecolor{view}{HTML}{FFFFFF}
\definecolor{bn}{HTML}{FFD700}
\tikzset{fc/.style={black,draw=black,fill=fc,rectangle,minimum height=1cm}}
\tikzset{h/.style={black,draw=black,fill=h,rectangle,minimum height=1cm}}
\tikzset{bias/.style={black,draw=black,fill=bias,rectangle,minimum height=1cm}}
\tikzset{noise/.style={black,draw=black,fill=noise,rectangle,minimum height=1cm}}
\tikzset{convd/.style={black,draw=black,fill=convd,rectangle,minimum height=1cm}}
\tikzset{convs/.style={black,draw=black,fill=convs,rectangle,minimum height=1cm}}
\tikzset{convf/.style={black,draw=black,fill=convf,rectangle,minimum height=1cm}}
\tikzset{pool/.style={black,draw=black,fill=pool,rectangle,minimum height=1cm}}
\tikzset{up/.style={black,draw=black,fill=up,rectangle,minimum height=1cm}}
\tikzset{view/.style={black,draw=black,fill=view,rectangle,minimum height=1cm}}
\tikzset{bn/.style={black,draw=black,fill=bn,rectangle,minimum height=1cm}}

\def\papertitle{Deep Embeddings for Robust User-Based \\ Amateur Vocal Percussion Classification}
\def\firstauthor{Alejandro Delgado}
\def\secondauthor{Emir Demirel}
\def\thirdauthor{Vinod Subramanian}
\def\fourthauthor{Charalampos Saitis}
\def\fifthauthor{Mark Sandler}

\newif\ifpdf
\ifx\pdfoutput\relax
\else
   \ifcase\pdfoutput
      \pdffalse
   \else
      \pdftrue
\fi

\ifpdf 
  \usepackage[pdftex,
    pdftitle={\papertitle},
    pdfauthor={\firstauthor, \secondauthor, \thirdauthor, \fourthauthor, \fifthauthor},
    bookmarksnumbered, 
    pdfstartview=XYZ 
   ]{hyperref}

  \usepackage[figure,table]{hypcap}

\else 
  \usepackage[dvips,
    bookmarksnumbered, 
    pdfstartview=XYZ 
  ]{hyperref}  

  \usepackage[figure,table]{hypcap}
\fi

\hypersetup{
    colorlinks,%
    citecolor=black,%
    filecolor=black,%
    linkcolor=black,%
    urlcolor=black
}

\title{\papertitle}

\oneauthor
   {Alejandro Delgado$^1{^2}$, Emir Demirel$^1$, Vinod Subramanian$^1$, Charalampos Saitis$^1$, and Mark Sandler$^1$} {$^1$ Queen Mary University of London $^2$ Luminary Roli Ltd. \\ %
     {\tt \href{mailto:a.delgadoluezas@qmul.ac.uk}{a.delgadoluezas@qmul.ac.uk}}}

\begin{document}
\capstartfalse
\maketitle
\capstarttrue
\begin{abstract}
Vocal Percussion Transcription (VPT) is concerned with the automatic detection and classification of vocal percussion sound events, allowing music creators and producers to sketch drum lines on the fly. Classifier algorithms in VPT systems learn best from small user-specific datasets, which usually restrict modelling to small input feature sets to avoid data overfitting. This study explores several deep supervised learning strategies to obtain informative feature sets for amateur vocal percussion classification. We evaluated the performance of these sets on regular vocal percussion classification tasks and compared them with several baseline approaches including feature selection methods and a speech recognition engine. These proposed learning models were supervised with several label sets containing information from four different levels of abstraction: instrument-level, syllable-level, phoneme-level, and boxeme-level. Results suggest that convolutional neural networks supervised with syllable-level annotations produced the most informative embeddings for classification, which can be used as input representations to fit classifiers with. Finally, we used back-propagation-based saliency maps to investigate the importance of different spectrogram regions for feature learning.
\end{abstract}

\section{Introduction}
\label{s1}

Vocal Percussion Transcription (VPT) is a relatively old subfield in Music Information Retrieval (MIR) that is concerned with the detection and classification of vocal percussion sound events, sitting just between monophonic music transcription and speech recognition. The goal of VPT is to transcribe vocal percussion sound events into typical drum instrument classes, usually including kick drums, snare drums, and hi-hats. From here on, we will use the term \textit{boxeme} (mix of "beatbox" and "phoneme") to refer to these vocal percussion sound events, as adopted in \cite{evain2021human}.

While VPT is a relatively specific field in MIR, its models and techniques are mostly shared across other disciplines in MIR and sound event detection. Some of these include musical instrument recognition \cite{solanki2019music}, music transcription \cite{roman2018end}, query by vocal imitation \cite{mehrabi2018similarity}, and anomalous sound detection \cite{nunes2021anomalous}. Likewise, vocal percussion datasets are also used in areas like music cognition \cite{mehrabi2019vocal} and to study inter-personal differences and trends in vocal imitation styles \cite{delgado2020spectral, delgado2021phoneme} among others.

Vocal percussion comes in two main modalities: beatbox and amateur vocal percussion. In \textit{beatbox}, boxemes are produced using numerous parts of the vocal tract, including ones that are not used in normal speech \cite{patil2017comparison}. This modality has its own universal set of techniques from which beatboxers base their own \cite{stowell2008characteristics}. In contrast, \textit{amateur vocal percussion} involves performers with little or no previous experience in vocal percussion, which usually includes most musicians and music producers. Due to this lack of training, boxemes are mostly speech-like and their articulation is much less consistent than in beatbox \cite{patil2017comparison}. Also, as amateur performers do not follow vocal percussion techniques, they usually decide to use their own particular set of boxemes that differ from those of other performers \cite{delgado2019new}.

The VPT process is usually composed of the onset detection and classification subtasks. \textit{Onset detection} deals with the prediction of the exact moments in the audio waveform where boxemes start, whereas \textit{classification} tries to assign each boxeme the correct associated drum instrument label (e.g. kick drums for /p/ boxemes). In this study, we focus exclusively on the classification process, leaving vocal percussion onset detection for future research.

Most recent works in amateur vocal percussion classification carry out the classification process in a user-based fashion, as this is known to improve classification performance compared to user-agnostic approaches \cite{hipke2014beatbox, ramires2018user, delgado2021learning}. In this way, users show the classifier their particular way of vocalising drum instruments (training set) so that the algorithm can recognise those boxemes in the future, usually within a beatbox-style improvisation (test set). Training sets are usually recorded either by reproducing a predictable beatbox-style phrase multiple times \cite{ramires2017automatic}, which is called \textit{fixed phrase} strategy, or by recording individual audio files containing same-class boxemes \cite{delgado2019new}, which is called \textit{isolated samples} strategy. Independently of the recording methodology, these user-specific training sets often contain less than a hundred boxemes. This \textit{data bottleneck} limits the amount of input audio features that classifiers can take for modelling so as to minimise their risk of overfitting. In consequence, amateur vocal percussion classifiers are in need of naturally informative input feature sets to guarantee robustness in prediction accuracy for all participants, irrespective of their stylistic idiosyncrasies and vocal percussion skills. As the informative power of a feature set depends largely on the task at hand, the exploration and evaluation of potentially informative feature sets would necessarily have to pass through boxeme analysis routines including heuristic feature selection and/or representation learning.

The present work explores the potential of several deep supervised learning approaches to generate informative feature sets for amateur vocal percussion classification. These feature learning strategies have been proven to be powerful feature extractors for high-dimensional data including sound events, music, and speech \cite{latif2020deep}. We supervise deep neural networks on four different types of label sets and take the values in their penultimate layer as the final feature sets to be evaluated. The four label sets that supervise the algorithms describe vocal percussion boxemes at different levels of abstraction, namely at instrument-level, syllable-level, phoneme-level, and boxeme-level. We assess the informative power of each of the learnt feature sets in terms of their classification accuracy and the stability of this to random train-validation splitting and initialisation routines. Finally, we carry out a complementary investigation of how relevant different regions in the spectrogram are for the models by applying saliency maps \cite{simonyan_deep_2013}.

\section{Related Work}
\label{s2}

\begin{table}
\centering
\setlength\tabcolsep{3.5pt}
\begin{tabular}{ccc}
& \mc{AVP Dataset} & \mc{LVT Dataset}\\
\midrule
Number of Participants  & 28 & 20\\
\addlinespace
Number of Boxemes  & 4,873  & 841\\
\addlinespace
Recording Strategy  & Isolated Samples  & Fixed Phrase\\
\addlinespace
Instrument Labels  &\textit{ kd, sd, hhc, hho}  & \textit{kd, sd, hhc}\\
\addlinespace
Phoneme Labels  & Yes  & Yes\\
\bottomrule
\end{tabular}
\caption{Summary of datasets' contents (kd = kick drum, sd = snare drum, hhc = closed hi-hat, hho = opened hi-hat).}
\label{t0}
\end{table}


The use of VPT algorithms was perhaps first applied to beatboxing via \cite{kapur2004query} and \cite{nakano2004drum}. These two studies feature three and two boxeme types respectively and both approach VPT by considering both the acoustic and rhythmic information contained in beatbox performances. In \cite{hazan2005towards}, Hazan proposed segmenting the boxemes in time prior to classification by applying an energy-based sound onset detector and explored K-Nearest Neighbours (KNN) and C4.5 tree algorithms for classification. This tree classifier was later revisited by Sinyor et al. \cite{sinyor2005beatbox}, which explicitly included amateur vocal percussion boxemes in the evaluation dataset. Stowell et al. \cite{stowell2010delayed} conducted experiments on inter-class separability of feature vectors from beatbox sound events and discovered that a better classification performance in the real-time regime could be achieved by applying a 23-milliseconds delay to the start of the classification analysis frame from the actual onset. All these studies used heuristic feature extraction \cite{peeters2011timbre} and traditional machine learning methods \cite{sammut2011encyclopedia} to train and evaluate boxeme classifiers.


Years later, novel approaches to VPT emerged inspired by recent advances in speech recognition, music information retrieval, and sound event classification techniques. In \cite{picart2015analysis}, Picart et al. explored audio pitch-tracking as a complement to the sound onset detection and classification engines, using a Hidden Markov Model (HMM) for transcription. This study also featured vocal imitations of pitched sounds from several musical instruments, where pitch-tracking algorithms proved the most useful. Ramires recorded in \cite{ramires2017automatic} the Live Vocalised Transcription (LVT) dataset, which was the first publicly available amateur vocal percussion dataset, and later developed the homonymous LVT system based on KNN vocal percussion classifiers \cite{ramires2018user}. In this study, we only use the acronym ``LVT'' when referring to the dataset. Delgado and colleagues \cite{delgado2019new} recorded another large publicly available amateur vocal percussion set, the Amateur Vocal Percussion (AVP) dataset, and carried out an evaluation of several onset detection algorithms. The same authors later explored data augmentation and deep learning techniques for user-based boxeme classification \cite{delgado2021learning}, where Convolutional Neural Network (CNN) models \cite{li2021survey} achieved the best accuracy results. Finally, Evain et al. \cite{evain2021human} adapted a popular HMM-based tool for automatic speech recognition \cite{povey2011kaldi} to VPT. The model was trained on a corpus of eighty different boxemes recorded individually by two beatboxers and yielded the best results using 22 Mel Frequency Cepstral Coefficients (MFCC) and their derivatives as the input features.

In the present study, we build on the works above and address several shortcomings of these, especially when it comes to final real-world implementations. These limitations include (i) the little availability of publicly available vocal percussion datasets, which impacts reproducibility, (ii) the low amount of participants and small size of the available datasets, which impacts the soundness and generalisability of results, (iii) the lack of beatbox-like improvisatory performances in some studies, which impacts the validity of results' extrapolation to real-world scenarios, and (iv) the lack of publicly available code, which impacts transparency. We tackle these issues respectively by (i) joining two publicly available datasets for amateur vocal percussion (AVP \cite{delgado2019new} and LVT \cite{ramires2017automatic}) and adapting the mix to VPT evaluation routines, (ii) using audio data augmentation techniques to improve external generalisability, (iii) testing final algorithms on freestyle improvisations to better assess their real-world capabilities, and (iv) publishing our code in an open-source repository\footnote{\textit{https://github.com/alejandrodl/vocal-percussion-transcription}}.

\section{Methodology}
\label{s3}

In this section, we provide an in-detail account of the main data sources, algorithms, and routines used throughout our study. In section \ref{s31}, we talk about the two datasets that we used (AVP and LVT), how we joined them and expanded their annotations so as to include phonetic information, and how we built input representations and carried out the data augmentation process. In section \ref{s32}, we describe the architecture of the embedding learning model and the seven types of label sets that we used for its supervision. In section \ref{s33}, we present the three baseline methods whose performances were compared to those of the embedding learning model supervised on different label sets and in section \ref{s34} we show how both the training and the evaluation processes were carried out.

\subsection{Data and Pre-Processing}
\label{s31}

We used two publicly available vocal percussion datasets throughout the study: the AVP dataset \cite{delgado2019new} and the LVT dataset \cite{ramires2017automatic}. We contrast some of these datasets' characteristics in Table \ref{t0}. The AVP dataset contains a total of 9778 boxemes (4873 and 4905 from the personal and fixed subsets respectively) recorded by 28 participants and with four annotated labels: kick drum, snare drum, closed hi-hat, and opened hi-hat. Its training dataset was recorded using the isolated samples strategy. To train our acoustic models, we exclusively used the personal subset (participants vocalising boxemes of their choice), although we also used the fixed subset (participants vocalising common boxemes) to train the sequential module of the baseline speech recognition model (see section \ref{s33}). The LVT dataset contains a total of 841 boxemes recorded by 20 participants with three annotated labels: kick drum, snare drum, and closed hi-hat. Here, the training dataset was recorded using the fixed phrase strategy. Also, we exclusively use its third subset, as the recordings' quality and background noise level are similar to those from the AVP dataset.



We manually expanded the annotations (onsets and instrument labels) of both datasets so as to include the syllabic representation of boxemes. The syllables were composed of a first \textit{onset phoneme}, usually plosive or fricative, and a second \textit{coda phoeneme}, usually a vowel, a breath sound, or silence (no coda phoneme). These phonemes were annotated following notation conventions from the International Phonetic Alphabet (IPA). Apart from the \textit{original} phoneme set, we also elaborated a \textit{reduced} phoneme set in which several similar-sounding phonemes were put together to form single classes. For this reduced version of phoneme annotations, onset and coda phonemes were grouped as shown in Table \ref{t1}. We also made the final AVP-LVT dataset with expanded annotations publicly available\footnote{\textit{https://zenodo.org/record/5578744\#.Yfpu9PXP30o}}.


\begin{table}[]
\centering
\begin{tabular}{|c|c|}
\hline
\textbf{Onset Phonemes}                                                                                              & \textbf{Coda Phonemes}                                                                                            \\ \hline
/t/ and /!/                                                                                                          & /\textscripta/, /\ae/, /\textturna/, and /\textturnv/ \\ \hline
/ts/ and /s/                                                                                                         & /e/, /\oe/, and /\textschwa/                                                        \\ \hline
/t\textesh/, /t\textctc/, /d\textyogh/, and /t\textyogh/ & /i/, /y/, and /\textsci/                                                                           \\ \hline
/kx/, /k/, and /k\textesh/                                                                            & /o/ and /\textupsilon/                                                                             \\ \hline
/p/ and /\textbarglotstop\textbarrevglotstop/                                          & /u/ and /\textturnm/                                                                               \\ \hline
\end{tabular}
\caption{Phoneme groupings for the reduced sets.}
\label{t1}
\end{table}

Joining the AVP and LVT datasets, we had a total of 5714 boxemes. As this amount of data was relatively modest for deep embedding modelling, we applied \textit{waveform data augmentation} to boxemes, specifically random \textit{pitch-shifting} (semitone range = [-1.5,+1.5]) and \textit{time-stretching} (stretch factor range = [0.8,1.2]), one after the other in random order. This kind of data augmentation is standard in audio signal processing \cite{nanni2020data} and it has been proven to improve the accuracy of vocal percussion classification algorithms \cite{delgado2021learning}. We applied ten iterations of random data augmentation in this manner, ending up with a total of 62854 boxemes in the final dataset.

As input to neural networks, we built Mel spectrogram representations from each boxeme using 64 Mel frequency bands and a hop size of 12 milliseconds. We used 48 time steps ($\sim$ 0.56 seconds) so that the final boxeme spectrograms had a final dimension of 64x48 and we explored frame sizes of 23, 46, and 93 milliseconds, ultimately reporting the one that brought the best results in Section \ref{s31}. We post-processed spectrograms with a logarithmic transform ($log(spectogram+0.0001)$) and normalised them to a [0,1] range.

\subsection{Supervised Embedding Learning}
\label{s32}

\begin{figure}
  \centering
  \resizebox{234pt}{180pt}{
  \begin{tikzpicture}

    \node[convd,rotate=90,minimum width=4.5cm] (conv1) at (2.5,0) {\large$\text{conv}_{1, 8}$\,+\,$\ReLU$\,+\,$\text{BN}$};
    \node[pool,rotate=90,minimum width=4.5cm] (pool1) at (3.75,0) {\large$\text{pool}_{2}$};
    \node[convd,rotate=90,minimum width=4.5cm] (conv2) at (5,0) {\large$\text{conv}_{8, 16}$\,+\,$\ReLU$\,+\,$\text{BN}$};
    \node[pool,rotate=90,minimum width=4.5cm] (pool2) at (6.25,0) {\large$\text{pool}_{2}$};
    \node[convd,rotate=90,minimum width=4.5cm] (conv3) at (7.5,0) {\large$\text{conv}_{16, 32}$\,+\,$\ReLU$\,+\,$\text{BN}$};
    \node[pool,rotate=90,minimum width=4.5cm] (pool3) at (8.75,0) {\large$\text{pool}_{2}$};
    \node[convd,rotate=90,minimum width=4.5cm] (conv4) at (10,0) {\large$\text{conv}_{32, 64}$\,+\,$\ReLU$\,+\,$\text{BN}$};
    \node[pool,rotate=90,minimum width=4.5cm] (pool4) at (11.25,0) {\large$\text{pool}_{2}$};

    \node[view,rotate=90,minimum width=4.5cm] (view4) at (12.5,0) {\large$\text{flatten}_{(64,4,4) \rightarrow 1024}$};
   
    \node[fc,rotate=90,minimum width=4.5cm] (fc1) at (13.75,0) {\large$\text{fc}_{1024, emb}$};
    \node[fc,rotate=90,minimum width=4.5cm] (fc2) at (15,0) {\large$\text{fc}_{emb, cls}$};
   
    \node[convd,rotate=90,minimum width=4.5cm] (conv1-2) at (2.5,-5) {\large$\text{conv}_{8, 8}$\,+\,$\ReLU$\,+\,$\text{BN}$};
    \node[convd,rotate=90,minimum width=4.5cm] (conv2-2) at (5,-5) {\large$\text{conv}_{16, 16}$\,+\,$\ReLU$\,+\,$\text{BN}$};
    \node[convd,rotate=90,minimum width=4.5cm] (conv3-2) at (7.5,-5) {\large$\text{conv}_{32, 32}$\,+\,$\ReLU$\,+\,$\text{BN}$};
    \node[convd,rotate=90,minimum width=4.5cm] (conv4-2) at (10,-5) {\large$\text{conv}_{64, 64}$\,+\,$\ReLU$\,+\,$\text{BN}$};
   
    \draw[->] (conv1) -- (conv1-2);
    \draw[->] (conv1-2) -- (pool1);
    \draw[->] (pool1) -- (conv2);
   
    \draw[->] (conv2) -- (conv2-2);
    \draw[->] (conv2-2) -- (pool2);
    \draw[->] (pool2) -- (conv3);
   
    \draw[->] (conv3) -- (conv3-2);
    \draw[->] (conv3-2) -- (pool3);
    \draw[->] (pool3) -- (conv4);
   
    \draw[->] (conv4) -- (conv4-2);
    \draw[->] (conv4-2) -- (pool4);
    \draw[->] (pool4) -- (view4);

    \draw[->] (view4) -- (fc1);
    \draw[->] (fc1) -- (fc2);

  \end{tikzpicture}}
  \caption{Diagram of the CNN embedding learning model. The variable $emb$ refers to the number of final embeddings to be extracted by removing the last layer and $cls$ refers to the number of classes with which the model is pretrained.}
  \label{f-1}
\end{figure}
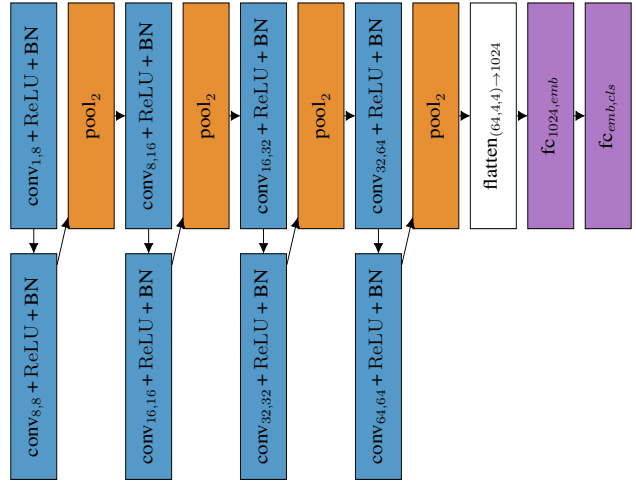

We used the penultimate layers of several CNN classifier models as the final embeddings to evaluate on. The architecture of these CNN models is illustrated in Figure \ref{f-1}. They all had four convolutional blocks with 8, 16, 32, and 64 filters respectively and two fully-connected (FC) linear layers, one connecting the flattened feature maps to the embedding space and another one connecting the embedding space to the labels. Each convolutional block had two convolutional layers with kernels of size 3x3 and stride 1x1, each one followed by a batch normalisation module and a ReLU activation gate. A final max-pooling operator with kernel size 2x2 is applied at the end of each convolutional block so as to progressively downsample the feature maps. The design of the network's architecture and its training routines (see section \ref{s34}) was inspired by best practices in CNN-based research \cite{he2019bag, santurkar2018does}.


We explored seven different supervision strategies to train and validate the above mentioned CNN classifiers. These strategies use different types of labels that describe the same input data at different levels of abstraction.

\subsubsection{Instrument-Level Annotations}
\label{s321}

We used two types of drum instrument annotations: the ones relative to the \textit{original} set (kick drum, snare drum, closed hi-hat, and opened hi-hat) and the ones relative to a \textit{reduced} version of it (drum and hi-hat). These constituted the first and the second supervision strategies.

\subsubsection{Syllable-Level Annotations}
\label{s322}

We also used syllable labels, put together by joining the onset and coda phonemes in the \textit{original} and the \textit{reduced} phoneme sets (see table \ref{t1}). These constituted the third and the fourth supervision strategies.

\subsubsection{Phoneme-Level Annotations}
\label{s323}

We used individual phoneme labels, which also came from the \textit{original} and the \textit{reduced} phoneme sets. These constituted the fifth and the sixth supervision strategies. It is worth noting that, while the two phoneme-level sets contained the same information as the two syllable-level ones, here the CNN classifier predicts onset and coda phoneme labels separately in a multi-task way, managing two different validation losses and accuracies.

\subsubsection{Boxeme-Level Annotations}
\label{s324}

Finally, we used boxeme labels to describe boxemes at the lowest level of abstraction. These boxeme classes were integrated by sounds that had different associated syllables and also pertained to different participants. This constitutes the seventh supervision strategy.



\subsection{Baseline Algorithms}
\label{s33}

\begin{figure*}
\centering
\resizebox{0.82\textwidth}{!}{
\begin{tikzpicture}[node distance=1.5cm, every node/.style={fill=white, font=\rmfamily}, align=center]


  \node (dataset)           [orange]                   {AVP-LVT Dataset (48)};
  \node (dataset_train) [orange, below of=dataset, xshift=-4.5cm] {Train-Validation Set (40)};
  \node (dataset_eval) [orange, below of=dataset, xshift=4.5cm] {Evaluation Set (8)};
 
  \node (data_augmentation) [orange, below of=dataset_train, yshift=-0.6cm] {Augmented Train-Validation};
  \node (trained_model) [green, below of=data_augmentation, yshift=-1cm] {Trained CNN Classifiers (25)};
  \node (embedding_model) [green, below of=trained_model, yshift=-0.6cm] {CNN Embedding Models (25)};
 
  \node (dataset_eval_train) [orange, below of=dataset_eval, xshift=-2.75cm] {Ev-Train Set};
  \node (dataset_eval_test) [orange, below of=dataset_eval, xshift=2.75cm] {Ev-Test Set};
 
  \node (data_augmentation_evtrain) [orange, below of=dataset_eval_train, yshift=-0.6cm] {Augmented Ev-Train};
  \node (embeddings_evtrain) [orange, below of=data_augmentation_evtrain, yshift=-0.6cm] {Embeddings Ev-Train (25)};
  \node (knn_classifier) [green, below of=embeddings_evtrain, yshift=-0.6cm] {Trained KNN Classifier (125)};
 
  \node (raw_accuracies) [red, below of=dataset_eval_test, yshift=-0.6cm] {Raw Accuracies (125)};
  \node (final_accuracies) [red, below of=raw_accuracies, yshift=-0.6cm] {Final Accuracy};
 
 
  \draw[->] (dataset) -- (dataset_train);
  \draw[->] (dataset) -- (dataset_eval);
 
  \draw[->] (dataset_eval) -- (dataset_eval_train);
  \draw[->] (dataset_eval) -- (dataset_eval_test);
 
  \draw[->] (dataset_train) -- node[text width=4cm] {10-fold data augmentation} (data_augmentation);
  \draw[->] (data_augmentation) -- node[text width=4cm] {5-fold CV \& 5 random initialisations} (trained_model);
  \draw[->] (trained_model) -- node[text width=4cm] {Remove last layer} (embedding_model);
 
  \draw[->] (dataset_eval_train) -- node[text width=4cm] {10-fold data augmentation} (data_augmentation_evtrain);
  \draw[->] (data_augmentation_evtrain) -- node[text width=4cm] (emb_cal) {Embeddings calculation} (embeddings_evtrain);
  \draw[->] (embeddings_evtrain) -- node[text width=4cm] {KNN with K=3,5,7,9,11} (knn_classifier);
 
  \draw[->] (dataset_eval_test) -- node[text width=4cm] (cls) {Classification} (raw_accuracies);
  \draw[->] (raw_accuracies) -- node[text width=4cm] {Mean and STD} (final_accuracies);
 
  \draw[->] (embedding_model) -- ++(3.2,0) -- ++(0,2.1) -- ++(1.2,0) (emb_cal);
  \draw[->] (knn_classifier) -- ++(2.9,0) -- ++(0,5.3) -- ++(1.48,0) (cls);

  \end{tikzpicture}}
 
\caption{Diagram of the training-evaluation process. Background colour code: orange = dataset-related, green = model-related, red = results-related.}
\label{f0}
\end{figure*}
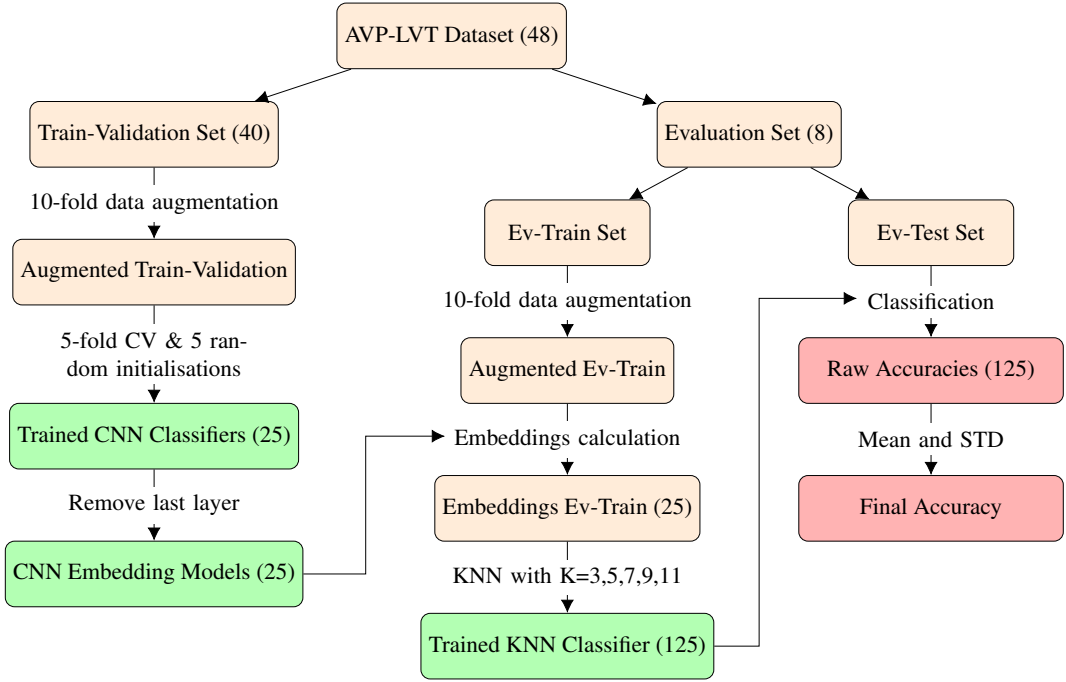

We compare the performance of learnt embeddings with two baseline heuristic feature sets and a speech recognition model. We used the Essentia toolbox \cite{bogdanov2013essentia} to calculate audio descriptors and the Kaldi library \cite{povey2011kaldi} to implement the speech recognition model.

\subsubsection{Timbre Feature Set}
\label{s331}

The first baseline model is the \textit{timbre feature set}, which is made of Mel Frequency Cepstral Coefficients (MFCC) and envelope features. The MFCCs are computed using the same frame-wise analysis parameters as for the spectrogram representations. For the 32-dimensional feature vector (see Section \ref{s34}), we take the mean of the first 14 MFCCs, the mean of their first derivative, and four envelope descriptors: the derivative after the maximum amplitude, the maximum derivative before the maximum amplitude, the flatness coefficient, and the temporal centroid to total length ratio. For the 16-dimensional feature vector, we take the mean of the first 12 MFCCs along with the four envelope descriptors.

\subsubsection{Feature Selection}
\label{s332}

For this method, we extracted 258 heuristic features from the spectrum and the envelope of boxemes and implemented seven importance-based \textit{feature selection} algorithms whose base algorithms were supervised using the same seven supervision strategies described in Section \ref{s32}. Here, instead of learning embeddings from spectrogram representations, we extracted the set of heuristic features, derived feature importances from a random forest base algorithm through a 10-way feature permutation process \cite{altmann2010permutation}, and selected the most informative features to build the final set. In the case of phoneme-level supervision strategies, the final selected features were drawn from the intersection of the two independent feature importances arrays, gathering the features that were considered most important to classify both onset and coda phonemes. We only reported the best result from these seven approaches in Section \ref{s4}.
 

\subsubsection{GMM-HMM Speech Recogniser}
\label{s333}

For the last baseline method, we tackled vocal percussion classification by means of \textit{speech recognition}, an approach proposed by Evain et al. in \cite{evain2021human} that was originally applied to beatbox classification with satisfactory results and that we bring to amateur vocal percussion classification in order to provide context for our methods. Here, each instrument type to detect is considered as a word to be recognized. First, an \textit{acoustic model} was trained to learn a relationship between acoustic speech features and phonemes. The mapping between instrument types and their corresponding phonemes was established through a pronunciation dictionary, converting phoneme posterior probabilities into word probabilities. We exploited both instrument- and phoneme-level annotations to construct this pronunciation dictionary. Word probabilities were then smoothed using a language model so as to obtain grammatically sensible transcriptions. The \textit{language model} was 5-gram trained on the transcriptions from training data, and the recogniser was trained via the Kaldi GMM-HMM recipe using the code in \textit{https://github.com/emirdemirel/ALTA} \cite{demirel2020automatic}. This was a triphone model trained with speaker adaptive features \cite{anastasakos1996compact}. Hyperparameters were determined empirically and they are available in the paper's repository.

\subsection{Training and Evaluation}
\label{s34}

A schematic diagram of the training and evaluation processes is provided in figure \ref{f0}.

We chose four participants from the AVP dataset and four from the LVT dataset to compose the evaluation set. Two women and two men per dataset were selected for evaluation on the basis of acceptable pronunciation and overall representativeness of the dataset. In the end, we had a total of 8 participants in the evaluation set and 40 participants in the train-validation set. We provide the distribution of the instrument, syllable, onset phoneme, and coda phoneme labels in the project's code repository.

We trained our seven CNN models (see sections \ref{s321}, \ref{s322}, \ref{s323}, and \ref{s324}) using the train-validation dataset. We set up a 5-fold cross-validation routine in which models were trained and validated for 5 iterations per fold, each with different initialisation parameters. Therefore, we end up with 25 deep embedding models per supervision strategy that encapsulate two main sources of training arbitrariness: the content of train-validation splits and random weight initialisation. We compute the mean and the 95\% confidence intervals of the subsequent 25 evaluation performances when reporting final results for a given supervision strategy. The same train-validation arrangement applies to the seven baseline feature selection methods. We trained the models using an Adam optimisation algorithm \cite{kingma2014adam}, early stopping if validation loss has not decreased after 10 epochs, and a further regularisation routine that downscales the learning rate if validation loss has not decreased after 5 epochs. In the case of phoneme-level supervision, we explored two settings of loss weights to compute the joint loss value after each training batch. The onset-coda weight percentages for those two settings were 50-50\% and 60-40\%.

As explained in section \ref{s31}, the 8 participants in the evaluation set have their own train and test subsets. From here on, we refer to these as the \textit{ev-train} and the \textit{ev-test} sets respectively to improve readability. During evaluation, a KNN algorithm was trained on the ev-train set of each participant and evaluated on the ev-test, taking the learnt CNN embeddings and the baseline feature sets as input representations. As these embeddings and feature sets are expected to have different information for the KNN algorithm to rely on, they are also expected to make the classifier perform better or worse given the underlying suitability of these feature sets for classification. As KNN algorithms are purely distance-based and non-parametric, our assumption is that a high classification accuracy using them is more likely to translate into a high accuracy using other types of machine learning algorithms. In Section \ref{s4}, we had the opportunity to briefly test this hypothesis by replacing the KNN classifier with three other popular machine learning classifiers whose results are commented on but not reported.

The evaluation procedure above is applied to all proposed and baseline methods except for the GMM-HMM-based speech recognition model. This model does not extract embeddings, but rather predicts the ev-test labels directly ignoring the ev-train subset (user-agnostic). Also, to alleviate the disparity in the amount of ev-train data in the AVP and the LVT datasets (1,000 vs. 220 augmented data samples approx.), we extract two different amounts of embeddings and selected features: 32 to carry out evaluation on AVP participants and 16 to do it on LVT ones. This means that for each supervision method we end up having 25 feature sets of size 32 to evaluate on AVP data and other 25 of size 16 to do it on LVT data.

\begin{table*}
\centering
\setlength\tabcolsep{1.41pt}
\begin{tabular}{cccccccccccc}
& \multicolumn{3}{c}{Baseline (3.3.\{1-2-3\})} & \multicolumn{2}{c}{Instrument (\ref{s321})} & \multicolumn{2}{c}{Syllable (\ref{s322})} & \multicolumn{2}{c}{Phoneme (\ref{s323})} & \multicolumn{1}{c}{Boxeme (\ref{s324})}\\
\cmidrule(l){2-4} \cmidrule(l){5-6} \cmidrule(l){7-8} \cmidrule(l){9-10} \cmidrule(lr){11-11}
& \mc{Timbre} & \mc{Selection} & \mc{HMM$^{*}$} & \mc{Original} & \mc{Reduced} & \mc{Original} & \mc{Reduced} & \mc{Original} & \mc{Reduced} & \mc{Original}\\
\midrule
Part-wise  & .840 & .827$\pm$.030 & .725 & .812$\pm$.037 & .779$\pm$.034 & \textbf{.899$\pm$.025} & .883$\pm$.030 & .876$\pm$.028 & .874$\pm$.030 & .861$\pm$.030\\
\addlinespace
Box-wise  & .835 & .795$\pm$.011 & .734 & .774$\pm$.038 & .738$\pm$.033 & \textbf{.874$\pm$.029} & .852$\pm$.031 & .840$\pm$.029 & .838$\pm$.032 & .832$\pm$.031\\
\bottomrule
\multicolumn{5}{l}{\footnotesize ${}^{*}$ \textit{User-agnostic model}}\\
\end{tabular}
\caption{Final evaluation accuracies from generated feature sets. Results are given participant-wise and boxeme-wise, and best performances for both modalities are highlighted in bold font. For feature selection, only the best performance is reported (reduced syllable-level). For embedding learning (25 models), the mean performances and their 95\% confidence intervals are reported.}
\label{t2}
\end{table*}

We report final results in two modalities: \textit{participant-wise}, where accuracy scores of single test participants are calculated and averaged, and \textit{boxeme-wise}, where accuracy scores are calculated for all evaluation boxemes in a participant-agnostic way.

Finally, we used \textit{saliency maps} \cite{simonyan_deep_2013} to highlight sections of the input that are important for model prediction. These maps can be computed given an input $x$ and the penultimate layer output of a deep learning model $A$ as: $\frac{dA_i}{dx}$, where $i$ is the label with respect to which the saliency map is computed. In our case, $A$ is the CNN model conditioned on the original instrument labels. We aggregated the saliency map by taking the absolute value and thresholding it so that the top 10\% of the map is set to one and the rest is set to zero. Then we averaged the saliency map over the 25 models.

\section{Results}
\label{s4}

\subsection{Classification}
\label{s41}

Final evaluation accuracies for all methods are gathered in Table \ref{t2}. Best results were obtained using a frame size of 46 ms and, in the case of phoneme-level classification, loss weights of 0.6 and 0.4 for onset and coda phonemes. Also, the best-performing feature selection routine was the one using the original syllable label set.


We can observe in the table that all supervised embedding models except for those supervised with instrument-level classes are consistently superior to baseline approaches, including feature selection approaches. This observation lies in accordance with previous literature on the usefulness of deep learning models as feature extractors for speech utterances \cite{latif2020deep}. It also highlights the unsuitability of instrument-level classes for embedding learning supervision, which was somewhat expected given that participants have different ways of vocalising drum instruments. Thus, label sets of such (high) level of abstraction are undesirable for embedding learning in our case.

We see that the best performance for both participant-wise and boxeme-wise evaluation metrics is achieved by supervised embedding learning models that used the original syllable-level label set. This difference is notable not only for the high mean accuracy score but also for its 95\% confidence intervals, which is the lowest for both participant-wise and boxeme-wise metrics. This means that the informative power of the resulting embeddings, apart from being the most prominent, is also the most robust to the two sources of training arbitrariness that we studied here (see Section \ref{s34}). We carried out experiments using other different classifiers than KNN, namely logistic regression, random forest, and extreme gradient boosted trees. There, we observed that the original syllable-level method still outperforms the rest of the approaches, which further reinforces its suitability for classification. All methods performed similarly on these extra algorithms, both in terms of absolute and relative performances.

Models supervised using original and reduced phoneme-level classes also yielded similar scores to those of the ones with syllable-level supervision, although still lower and generally less robust to training arbitrariness, possibly due to the extra complexity of the multi-task learning approach. The same applied to boxeme-level supervision, which performed slightly worse than phoneme-level supervision, possibly indicating that very low levels of supervision abstraction are relatively counterproductive for vocal percussion classification engines.

Another reason that could explain this lower performance for boxeme-level supervision could be its large amount of output labels ($\sim$150 boxeme types), which complicates validation. In order to tackle this issue, we built and trained an alternative siamese network model \cite{koch2015siamese} with the same architecture as the CNN except for the last fully connected layer, which was removed. This network uses metric learning directly on embeddings to discriminate between same-class boxeme pairs and different-class ones, therefore reducing our large label vector to a ``same-different'' binary auxiliary vector. In the end, its final accuracy was found to be moderately lower than the one relative to the CNN classifier, so we kept the latter's result.

We also notice that the generic (user-agnostic) HMM-based speech recognition model performs worse than any other user-based method. This result evidences the importance of taking user idiosyncrasies into account in amateur vocal percussion classification, making user-based strategies preferable to generic user-agnostic ones. Finally, we observe that accuracies derived using the timbre feature set are higher than all the ones pertaining to baseline feature selection algorithms. This result could indicate an excessive information redundancy of selected features compared to that of the timbre set, which is a more internally cohesive feature set.

A clear limitation of the syllable-level embeddings as inputs to amateur vocal percussion classifiers is that these were learnt using samples recorded with electronic devices in a small room with little background noise, which emulates the typical recording setting of amateur music producers. This is likely to work similarly well in recording studios, where audio quality is higher, but may very well lose a significant part of its accuracy when faced with low-quality recordings or contexts with too much noise. The timbre feature set could be of great help for these situations, as its features are context-agnostic and therefore adapt well to challenging recording scenarios.

\subsection{Saliency Maps}
\label{s42}

\begin{figure}
    \centering
    \subfloat[Kick drum]{\includegraphics[width=0.25\textwidth]{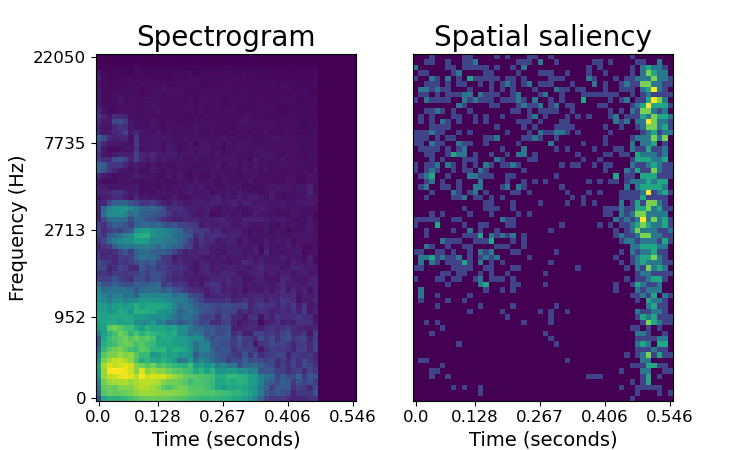}}
    \subfloat[Snare drum]{\includegraphics[width=0.25\textwidth]{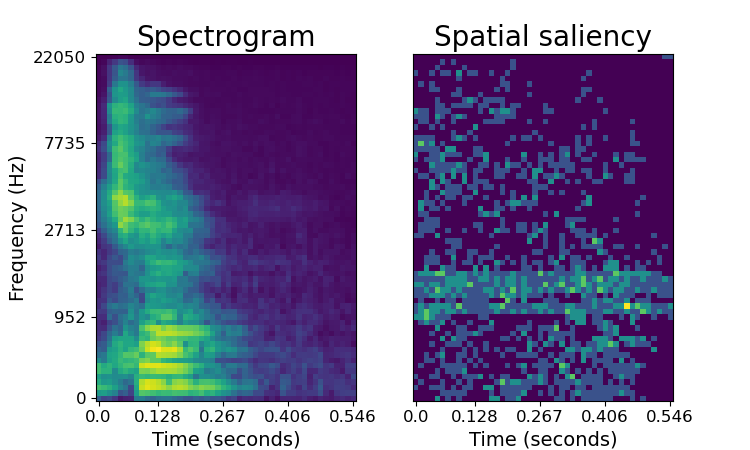}}\\
    \subfloat[Closed Hi-Hat]{\includegraphics[width=0.25\textwidth]{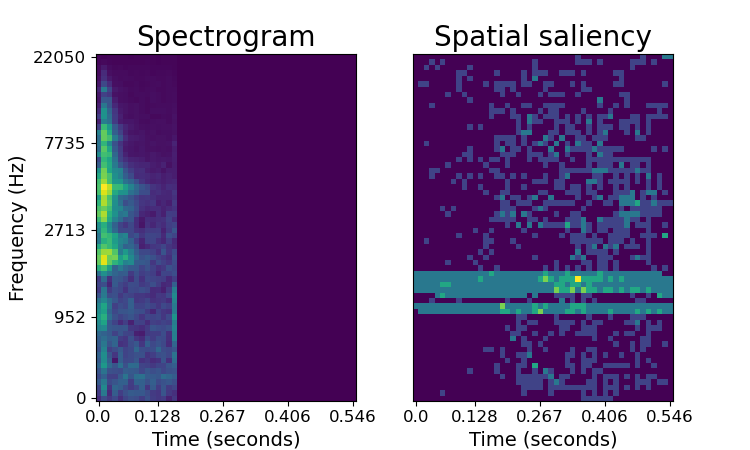}}
    \subfloat[Opened Hi-Hat]{\includegraphics[width=0.25\textwidth]{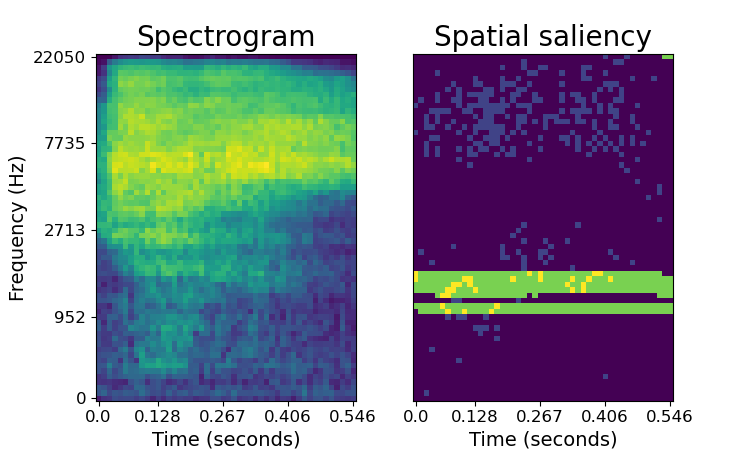}}
    \caption{Log mel spectrogram (left) and corresponding saliency map (right) of four boxemes of different instrument class.}
    \label{fig:aggregate_saliency}
\end{figure}

Four representative examples\footnote{See project's repository for more examples of saliency maps.} of these maps are shown in Figure \ref{fig:aggregate_saliency}. In general, we found that models tended to focus more on frequencies between $1000Hz$ and $2000Hz$ for boxemes associated with the snare drum, closed hi-hat, and opened hi-hat. This region usually coincides with a high-energy one for these boxemes, which often share phonetic representations. It also might indicate a useful thresholding point for the network to tell the boxemes apart. Hence, this could be implying that a higher frequency resolution in this region could potentially improve the accuracy of future amateur vocal percussion classifiers.

The network also appears to attend to silences and regions of lower spectral energy in the case of the kick drum and closed hi-hat. This could mean the absence of energy, especially at the high end of the spectrum, might also be a key feature for the models to differentiate the kick drum and closed hi-hat from the rest of the instruments. The high attention density in silences specifically could also be implicitly suggesting that the duration of boxemes is a key factor to distinguish the kick drum and closed hi-hat from the snare drum and opened hi-hat, as the boxemes associated with these last two instruments tend to be longer.


\section{Conclusion}
\label{s5}

We have explored the capabilities of supervised embedding models to generate informative feature sets that can achieve amateur vocal percussion classification in a fast and reliable way. We used seven supervision strategies whose label sets describe vocal percussion sound events (boxemes) at different levels of abstraction and we found that CNN classifiers supervised using syllable-level classes not only produced the best mean accuracy results but also were the most robust to sources of training randomness like weights initialisation. As discussed in the results section, we highly encourage the use of these embedding models as feature extractors for amateur vocal percussion classifiers in recording settings with low and moderate background noise. We have made both the AVP-LVT dataset and the code for this study publicly available.

\begin{acknowledgments}
This work has received funding from the European Union’s Horizon 2020 research and innovation programme under the Marie Skłodowska-Curie grant agreement No. 765068.
\end{acknowledgments}

\bibliography{smc2022bib}

\begin{thebibliography}{10}
\providecommand{\url}[1]{#1}
\csname url@samestyle\endcsname
\providecommand{\newblock}{\relax}
\providecommand{\bibinfo}[2]{#2}
\providecommand{\BIBentrySTDinterwordspacing}{\spaceskip=0pt\relax}
\providecommand{\BIBentryALTinterwordstretchfactor}{4}
\providecommand{\BIBentryALTinterwordspacing}{\spaceskip=\fontdimen2\font plus
\BIBentryALTinterwordstretchfactor\fontdimen3\font minus
  \fontdimen4\font\relax}
\providecommand{\BIBforeignlanguage}[2]{{%
\expandafter\ifx\csname l@#1\endcsname\relax
\typeout{** WARNING: IEEEtran.bst: No hyphenation pattern has been}%
\typeout{** loaded for the language `#1'. Using the pattern for}%
\typeout{** the default language instead.}%
\else
\language=\csname l@#1\endcsname
\fi
#2}}
\providecommand{\BIBdecl}{\relax}
\BIBdecl

\bibitem{evain2021human}
S.~Evain, B.~Lecouteux, D.~Schwab, A.~Contesse, A.~Pinchaud, and N.~H.
  Bernardoni, ``Human beatbox sound recognition using an automatic speech
  recognition toolkit,'' \emph{Biomedical Signal Processing and Control},
  vol.~67, p. 102468, 2021.

\bibitem{solanki2019music}
A.~Solanki and S.~Pandey, ``Music instrument recognition using deep
  convolutional neural networks,'' \emph{International Journal of Information
  Technology}, pp. 1--10, 2019.

\bibitem{roman2018end}
M.~A. Rom{\'a}n, A.~Pertusa, and J.~Calvo-Zaragoza, ``An end-to-end framework
  for audio-to-score music transcription on monophonic excerpts.'' in
  \emph{ISMIR}, 2018, pp. 34--41.

\bibitem{mehrabi2018similarity}
A.~Mehrabi, K.~Choi, S.~Dixon, and M.~Sandler, ``Similarity measures for
  vocal-based drum sample retrieval using deep convolutional auto-encoders,''
  in \emph{2018 IEEE International Conference on Acoustics, Speech and Signal
  Processing (ICASSP)}.\hskip 1em plus 0.5em minus 0.4em\relax IEEE, 2018, pp.
  356--360.

\bibitem{nunes2021anomalous}
E.~C. Nunes, ``Anomalous sound detection with machine learning: A systematic
  review,'' \emph{arXiv preprint arXiv:2102.07820}, 2021.

\bibitem{mehrabi2019vocal}
A.~Mehrabi, S.~Dixon, and M.~Sandler, ``Vocal imitation of percussion sounds:
  On the perceptual similarity between imitations and imitated sounds,''
  \emph{Plos one}, vol.~14, no.~7, p. e0219955, 2019.

\bibitem{delgado2020spectral}
A.~Delgado~Luezas, C.~Saitis, M.~Sandler \emph{et~al.}, ``Spectral and temporal
  timbral cues of vocal imitations of drum sounds.''\hskip 1em plus 0.5em minus
  0.4em\relax International Conference on Timbre, 2020.

\bibitem{delgado2021phoneme}
A.~Delgado, C.~Saitis, and M.~Sandler, ``Phoneme mappings for online vocal
  percussion transcription,'' in \emph{Audio Engineering Society Convention
  151}.\hskip 1em plus 0.5em minus 0.4em\relax Audio Engineering Society, 2021.

\bibitem{patil2017comparison}
N.~Patil, T.~Greer, R.~Blaylock, and S.~S. Narayanan, ``Comparison of basic
  beatboxing articulations between expert and novice artists using real-time
  magnetic resonance imaging.'' in \emph{Interspeech}, 2017, pp. 2277--2281.

\bibitem{stowell2008characteristics}
D.~Stowell and M.~D. Plumbley, ``Characteristics of the beatboxing vocal
  style,'' \emph{Dept. of Electronic Engineering, Queen Mary, University of
  London, Technical Report, Centre for Digital Music C4DMTR-08-01}, 2008.

\bibitem{delgado2019new}
A.~Delgado, S.~McDonald, N.~Xu, and M.~Sandler, ``A new dataset for amateur
  vocal percussion analysis,'' in \emph{Proceedings of the 14th International
  Audio Mostly Conference: A Journey in Sound}, 2019, pp. 17--23.

\bibitem{hipke2014beatbox}
K.~Hipke, M.~Toomim, R.~Fiebrink, and J.~Fogarty, ``Beatbox: End-user
  interactive definition and training of recognizers for percussive
  vocalizations,'' in \emph{Proceedings of the 2014 International Working
  Conference on Advanced Visual Interfaces}, 2014, pp. 121--124.

\bibitem{ramires2018user}
A.~Ramires, R.~Penha, and M.~E. Davies, ``User specific adaptation in automatic
  transcription of vocalised percussion,'' \emph{arXiv preprint
  arXiv:1811.02406}, 2018.

\bibitem{delgado2021learning}
A.~Delgado, S.~McDonald, N.~Xu, C.~Saitis, and M.~Sandler, ``Learning models
  for query by vocal percussion: A comparative study,'' in \emph{Proceedings of
  the International Computer Music Conference}, 2021.

\bibitem{ramires2017automatic}
A.~F.~S. Ramires, ``Automatic transcription of vocalized percussion,'' 2017.

\bibitem{latif2020deep}
S.~Latif, R.~Rana, S.~Khalifa, R.~Jurdak, J.~Qadir, and B.~W. Schuller, ``Deep
  representation learning in speech processing: Challenges, recent advances,
  and future trends,'' \emph{arXiv preprint arXiv:2001.00378}, 2020.

\bibitem{simonyan_deep_2013}
\BIBentryALTinterwordspacing
K.~Simonyan, A.~Vedaldi, and A.~Zisserman, ``Deep inside convolutional
  networks: Visualising image classification models and saliency maps.''
  [Online]. Available: \url{http://arxiv.org/abs/1312.6034}
\BIBentrySTDinterwordspacing

\bibitem{kapur2004query}
A.~Kapur, M.~Benning, and G.~Tzanetakis, ``Query-by-beat-boxing: Music
  retrieval for the dj,'' in \emph{Proceedings of the International Conference
  on Music Information Retrieval}, 2004, pp. 170--177.

\bibitem{nakano2004drum}
T.~Nakano, J.~Ogata, M.~Goto, and Y.~Hiraga, ``A drum pattern retrieval method
  by voice percussion,'' \emph{Database (Musical Instrument Sound)}, vol.~3,
  p.~1, 2004.

\bibitem{hazan2005towards}
A.~Hazan, ``Towards automatic transcription of expressive oral percussive
  performances,'' in \emph{Proceedings of the 10th international conference on
  Intelligent user interfaces}, 2005, pp. 296--298.

\bibitem{sinyor2005beatbox}
E.~Sinyor, C.~M. Rebecca, D.~Mcennis, and I.~Fujinaga, ``Beatbox classification
  using ace,'' in \emph{Proceedings of the International Conference on Music
  Information Retrieval}.\hskip 1em plus 0.5em minus 0.4em\relax Citeseer,
  2005.

\bibitem{stowell2010delayed}
D.~Stowell and M.~D. Plumbley, ``Delayed decision-making in real-time beatbox
  percussion classification,'' \emph{Journal of New Music Research}, vol.~39,
  no.~3, pp. 203--213, 2010.

\bibitem{peeters2011timbre}
G.~Peeters, B.~L. Giordano, P.~Susini, N.~Misdariis, and S.~McAdams, ``The
  timbre toolbox: Extracting audio descriptors from musical signals,''
  \emph{The Journal of the Acoustical Society of America}, vol. 130, no.~5, pp.
  2902--2916, 2011.

\bibitem{sammut2011encyclopedia}
C.~Sammut and G.~I. Webb, \emph{Encyclopedia of machine learning}.\hskip 1em
  plus 0.5em minus 0.4em\relax Springer Science \& Business Media, 2011.

\bibitem{picart2015analysis}
B.~Picart, S.~Brognaux, and S.~Dupont, ``Analysis and automatic recognition of
  human beatbox sounds: A comparative study,'' in \emph{2015 IEEE international
  conference on acoustics, speech and signal processing (ICASSP)}.\hskip 1em
  plus 0.5em minus 0.4em\relax IEEE, 2015, pp. 4255--4259.

\bibitem{li2021survey}
Z.~Li, F.~Liu, W.~Yang, S.~Peng, and J.~Zhou, ``A survey of convolutional
  neural networks: analysis, applications, and prospects,'' \emph{IEEE
  Transactions on Neural Networks and Learning Systems}, 2021.

\bibitem{povey2011kaldi}
D.~Povey, A.~Ghoshal, G.~Boulianne, L.~Burget, O.~Glembek, N.~Goel,
  M.~Hannemann, P.~Motlicek, Y.~Qian, P.~Schwarz \emph{et~al.}, ``The kaldi
  speech recognition toolkit,'' in \emph{IEEE 2011 workshop on automatic speech
  recognition and understanding}, no. CONF.\hskip 1em plus 0.5em minus
  0.4em\relax IEEE Signal Processing Society, 2011.

\bibitem{nanni2020data}
L.~Nanni, G.~Maguolo, and M.~Paci, ``Data augmentation approaches for improving
  animal audio classification,'' \emph{Ecological Informatics}, vol.~57, p.
  101084, 2020.

\bibitem{he2019bag}
T.~He, Z.~Zhang, H.~Zhang, Z.~Zhang, J.~Xie, and M.~Li, ``Bag of tricks for
  image classification with convolutional neural networks,'' in
  \emph{Proceedings of the IEEE/CVF Conference on Computer Vision and Pattern
  Recognition}, 2019, pp. 558--567.

\bibitem{santurkar2018does}
S.~Santurkar, D.~Tsipras, A.~Ilyas, and A.~Madry, ``How does batch
  normalization help optimization?'' \emph{Advances in neural information
  processing systems}, vol.~31, 2018.

\bibitem{bogdanov2013essentia}
D.~Bogdanov, N.~Wack, E.~G{\'o}mez~Guti{\'e}rrez, S.~Gulati, H.~Boyer,
  O.~Mayor, G.~Roma~Trepat, J.~Salamon, J.~R. Zapata~Gonz{\'a}lez, X.~Serra
  \emph{et~al.}, ``Essentia: An audio analysis library for music information
  retrieval,'' in \emph{Britto A, Gouyon F, Dixon S, editors. 14th Conference
  of the International Society for Music Information Retrieval (ISMIR); 2013
  Nov 4-8; Curitiba, Brazil.[place unknown]: ISMIR; 2013. p. 493-8.}\hskip 1em
  plus 0.5em minus 0.4em\relax International Society for Music Information
  Retrieval (ISMIR), 2013.

\bibitem{altmann2010permutation}
A.~Altmann, L.~Tolo{\c{s}}i, O.~Sander, and T.~Lengauer, ``Permutation
  importance: a corrected feature importance measure,'' \emph{Bioinformatics},
  vol.~26, no.~10, pp. 1340--1347, 2010.

\bibitem{demirel2020automatic}
E.~Demirel, S.~Ahlb{\"a}ck, and S.~Dixon, ``Automatic lyrics transcription
  using dilated convolutional neural networks with self-attention,'' in
  \emph{2020 International Joint Conference on Neural Networks (IJCNN)}.\hskip
  1em plus 0.5em minus 0.4em\relax IEEE, 2020.

\bibitem{anastasakos1996compact}
T.~Anastasakos, J.~McDonough, R.~Schwartz, and J.~Makhoul, ``A compact model
  for speaker-adaptive training,'' in \emph{Proceedings of the Fourth
  International Conference on Spoken Language Processing}.\hskip 1em plus 0.5em
  minus 0.4em\relax IEEE, 1996.

\bibitem{kingma2014adam}
D.~P. Kingma and J.~Ba, ``Adam: A method for stochastic optimization,''
  \emph{arXiv preprint arXiv:1412.6980}, 2014.

\bibitem{koch2015siamese}
G.~Koch, R.~Zemel, R.~Salakhutdinov \emph{et~al.}, ``Siamese neural networks
  for one-shot image recognition,'' in \emph{ICML deep learning workshop},
  vol.~2.\hskip 1em plus 0.5em minus 0.4em\relax Lille, 2015.

\end{thebibliography}

\end{document}